\documentclass[twocolumn,a4paper,10pt]{article}
\usepackage{amsmath,amsfonts}
\usepackage{epsfig}
\usepackage{times}
\usepackage[small,hang]{caption}
\usepackage{xcolor}
\usepackage[percent]{overpic}

\usepackage[style=authoryear]{biblatex}
\usepackage[normalem]{ulem}
\usepackage{color}
\usepackage{hyperref}
\usepackage{breakurl} 

\makeatletter  

\newcounter{numbersec}
\renewcommand{\section}[1]{\par\noindent\stepcounter{numbersec}
	\par
	\vspace{6pt}
	\noindent\textbf{\large   \arabic{numbersec} \hspace*{0.3cm} #1 }
	\par
	\vspace{2pt}
}
\renewcommand{\subsection}[1]{
	\par
	\vspace{6pt}
	\noindent\textbf{#1}
	\par
}
\renewcommand{\subsubsection}[1]{%
	\par
	\vspace{6pt}
	\textbf{#1.}
}

\newcommand{\Abstract}{\par\vspace{6pt}\noindent\textbf{\large Abstract}\par\vspace{2pt}}
\newcommand{\Acknowledgments}{\par\vspace{6pt}\noindent\textbf{\large Acknowledgments }\par\vspace{2pt}}

\newenvironment{References}{
\par\vspace{6pt}\noindent\textbf{\large References}\par\vspace{2pt}
\begin{small}\begin{list}{ }{\itemsep2mm \parsep0mm\labelsep0mm\leftmargin0mm}}
{\end{list}\end{small}}

\makeatother

\title{\vspace*{-12mm}
\LARGE \sc \textbf{  
Drag-reduction strategies in wall-bounded turbulent flows using deep reinforcement learning
}}

\author{ \Large \bf \textit{ 
L. Guastoni$^{1,2*}$, J. Rabault$^{3}$, H. Azizpour$^{4,2}$ and R. Vinuesa$^{1,2}$}  \\ \\
\normalsize \bf  \textit{$^{1}$FLOW, Engineering Mechanics, KTH Royal Inst. of Tech., Stockholm, Sweden} \\
\normalsize \bf  \textit{$^{2}$Swedish e-Science Research Centre (SeRC), Stockholm, Sweden} \\ 
\normalsize \bf  \textit{$^{3}$Independent researcher, Oslo, Norway} \\
\normalsize \bf  \textit{$^{4}$Div. of Robotics, Perception, and Learning, KTH Royal Inst. of
Tech., Stockholm, Sweden} \\
\underline{\normalsize \bf $^{*}$guastoni@mech.kth.se}
}
\date{}

\begin{document}
\maketitle
\thispagestyle{empty}

%%%%%%%%%%%%%%%%%%%%
% Paper text
%%%%%%%%%%%%%%%%%%%%

%%% Insert here the abstract %%%

\Abstract

In this work we compare different drag-reduction strategies that compute their actuation based on the fluctuations at a given wall-normal location in turbulent open channel flow. In order to perform this study, we implement and describe in detail the reinforcement-learning interface to a computationally-efficient, parallelized, high-fidelity solver for fluid-flow simulations.

We consider opposition control~(\cite{opposition}) and the policies learnt using deep reinforcement learning (DRL) based on the state of the flow at two inner-scaled locations ($y^+ = 10$ and $y^+ = 15$). 
 %By retrofitting an existing validated solver, we are able to test existing deep-reinforcement-learning (DRL) algorithms on a turbulence-control task, and discover drag-reduction control strategies that outperform existing control methods. 
% The approach used for the retrofitting can be extended to other high-performance computing codes, which can be too difficult or costly to rewrite or modify for new emerging research needs. 
By using deep deterministic policy gradient (DDPG) algorithm, we are able to discover control strategies that outperform existing control methods. This represents a first step in the exploration of the capability of DRL algorithm to discover effective drag-reduction policies using information from different locations in the flow. 

%%%  Insert here the actual article text %%% 

\section{Introduction} %%%%%%%%%%%%%%%%%%
The control of turbulent flows has high relevance, for both economical and environmental reasons. Whether the goal is to promote turbulence (\textit{e.g.} to enhance mixing) or reduce it, some form of flow control needs to be designed and deployed. Due to the complexity and cost of experimental studies, computational fluid dynamics (CFD) has been an important research tool in flow control studies. The use of numerical simulation allows researchers to test very fine-grained control strategies that are difficult or impossible to replicate in a realistic scenarios, but that can provide an insight on the flow features that can be targeted using the control. One example is  
% Traditional approaches rely on 
%reactive flow control, where the actuation is applied in order to prevent the development of specific flow structures. In 
\textit{opposition control}~(\cite{opposition,opposition2}): the wall-normal velocity fluctuations distribution at a given wall-normal location is sampled and then it is imposed at the wall with opposite sign. This approach aims to suppress the near-wall turbulent structures.

More recently, machine learning models have also been used to develop new flow control strategies, for instance,~\cite{park_choi_2020} used convolutional neural networks to compute the wall-normal velocity distribution to be applied at the wall, with the goal to reduce the turbulent drag in the flow. Another data-driven set of algorithms that is used to solve control tasks is reinforcement learning (RL). In particular, deep reinforcement learning is a mathematical framework that leverages neural networks to solve Markov decision processes (MDPs), which has been successfully applied to a range of fluid-dynamics problems.
Among the possible applications, we mention mesh optimization~(\cite{lorsung2023}), the maximization of the efficiency of agents swimming in a turbulent flow~(\cite{biferale,swimmers}) and turbulence modelling~(\cite{novati,kurz-modelling}).
In this work, we focus on a drag reduction problem: the recent advancements on the use of DRL for this task are summarized by~\cite{vignon}. The control of the two-dimensional wake behind a cylinder has been investigated in a number of research studies~(\cite{rabault, varela}). Very recently, three-dimensional and fully-turbulent drag-reduction problems have started to be analyzed with the same tools~(\cite{sonoda,zeng}). 
In this work we introduce a RL environment to simulate a turbulent flow in an open channel, using a multi-agent approach that allows the resulting control strategy to be local and translationally-invariant. The use of a multi-agent approach has been confirmed to be essential for the success of the learning procedure in other flow control applications such as reducing convective heat transfer in two-dimensional Rayleigh--B\'enard convection~(\cite{Vignon_2023}). A single agent is not able to effectively modify the heat transfer, whereas a set of agents that share the same policy is able to manipulate the flow configuration in a non-trivial way to reduce the Nusselt number $Nu$.

\section{Methodology}
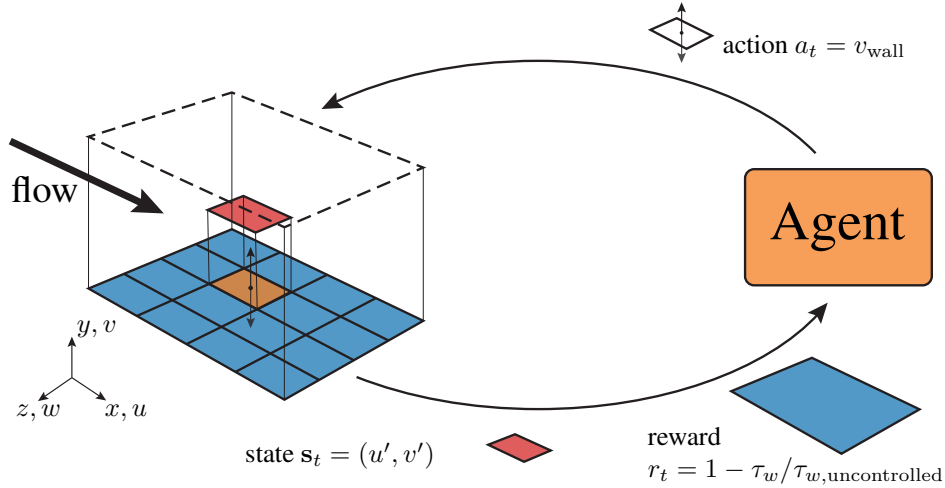
\begin{figure*}[t!]
\begin{center}
% \begin{overpic}[width=0.7\textwidth,grid,tics=10]{channel_flow_domain_nolabels.eps}
\begin{overpic}[width=0.77\textwidth]{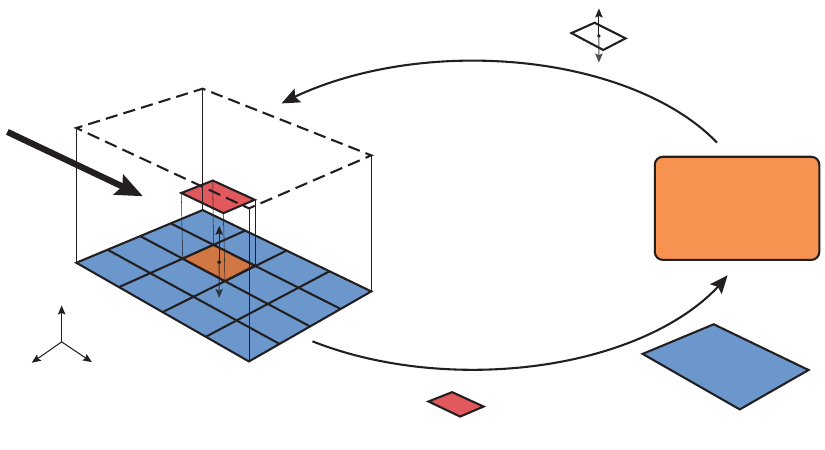}%[grid,tics=10]
 \put (1,32) {\Large flow}
 \put (82,28) {\huge Agent}
 \put (69,4) {\begin{minipage}{0.3\textwidth}reward\\ $r_t = 1-\tau_w/\tau_{w,\mathrm{uncontrolled}}$\end{minipage}}
 \put (26,4) {state $\mathbf{s}_t=(u',v')$}
 \put (77,48) {action $a_t=v_\mathrm{wall}$}
 \put (8,18) {$y,v$}
 \put (11,9) {$x,u$}
 \put (1.5,9) {$z,w$}
\end{overpic}
\end{center}
\caption{\label{teaser} Summary of our multi-agent DRL methodology to learn drag-reducing policies in turbulent channel flows. The computational domain is shown on the left. $N_{\rm{CTRLx}} \times N_{\rm{CTRLz}}$ agents at the wall are used to modify the flow. Each agents observes the velocity fluctuations in the streamwise ($u'$) and wall-normal ($v'$) direction. The reward variation of the wall-shear stress $\tau_w$ with respect to the uncontrolled case. The action of the agent is defined as a wall-normal velocity $v$ at the wall.}
\end{figure*}

Our implementation is summarized in figure~\ref{teaser}. %In the left part, we show the simulation domain for the open channel flow. is simulated. 
First, we describe the simulation setting in the left part of the figure. We perform a simulation of an open channel flow in a rectangular domain. 
We indicate the streamwise, wall-normal and spanwise directions with $(x,y,z)$, respectively and their corresponding velocity components by $(u,v,w)$.
% Periodic boundary conditions are imposed in the streamwise and spanwise directions, while a no-slip condition and a symmetry conditions are imposed at the two wall-normal boundaries.
We consider two computational domains, a minimal channel~(\cite{jimenez-minimal}), with size $\Omega = L_x \times L_y \times L_z = 2.67h \times h \times 0.8h$ (where $h$ is the open-channel height) and a larger channel with size $\Omega = L_x \times L_y \times L_z = 2\pi h \times h \times \pi h$. The resolution in the two DNSs is $N_x \times N_y \times N_z = 16 \times 65 \times 16$ and $N_x \times N_y \times N_z = 64 \times 65 \times 64$, respectively. The uncontrolled flow has a friction Reynolds number $Re_{\tau} = u_{\tau}h/\nu = 180$, where the friction velocity $u_{\tau}=\sqrt{\tau_w/\rho}$ (based on the the wall-shear stress $\tau_w$ and the fluid density $\rho$). $\nu$ is kinematic viscosity of the fluid. This value is the same used in previous drag-reduction studies~(\cite{opposition}).
%, which represents the minimal flow unit in which a self-sustained turbulent flow is possible. 
% A larger channel is also considered, with a different domain $\Omega$ and resolution $N_x \times N_y \times N_z$. Further details are available in appendix~\ref{appA}.
% The Reynolds number used in this study and in the fluid-dynamics literature for wall-bounded flows is the friction Reynolds number, defined as $Re_{\tau} = u_{\tau}h/\nu$, where the friction velocity $u_{\tau}=\sqrt{\tau_w/\rho}$ (based on the the wall-shear stress $\tau_w$ and the fluid density $\rho$). $\nu$ is kinematic viscosity of the fluid. We consider $Re_{\tau}=180$, to compare the drag reduction with the corresponding results in previous active flow control works~\citep{opposition}.

In order to frame our fluid-dynamics problem as a reinforcement learning problem, three elements need to be defined: the action performed by the agent, the observation of the environment that is used to compute the action and the reward obtained once the action has been performed. 
In our case, the chosen action is the application a wall-normal velocity at the wall, that allows the agent to alter the flow to reduce the drag.
One possible, naive, reinforcement-learning approach would be to let an agent select the wall-normal velocity in all the points. %The application of the control in each point of the wall boundary allows for the maximum control authority on all the simulated time- and length-scales in the flow. 
However, the resulting action space would be very large and very difficult to explore it in an efficient and comprehensive way. Furthermore, the learnt policy would not take advantage of the invariance by translation of the system. This latter feature can have a decisive impact on the success of the learning process, as highlighted by~\cite{belus2019exploiting}.
In order to address these issues, we opted for a multi-agent reinforcement-learning (MARL) approach: several agents that share the same policy are interacting in the same environment. We consider a grid of $N_{\rm{CTRLx}} \times N_{\rm{CTRLz}} = N_x \times N_z$ agents in the streamwise ($x$) and spanwise ($z$) directions, respectively. 
This approach allows to reduce the action space dimensionality and also enforce the translational invariance of the policy.

In our work, we define the state, action, and reward signals to be comparable with opposition control. The control is computed based on the velocity fluctuations at a given distance from the wall, above the area where it is selecting the actuation. Note that opposition control typically relies on the observation of the wall-normal component of the velocity fluctuations. In addition to this component, we also provide the streamwise velocity fluctuations.
The choice of the sampling height for the state observations are determine which kind of flow features the DRL agent can use to characterize the entire flow and control it. In~\cite{Guastoni2023}, the sampling height used in the numerical experiments is $y^+ = 15$, where $y^+$ indicated the inner-scaled coordinates using the viscous length $\ell^{*} = \nu/u_{\tau}$. This value is found to provide the highest drag reduction in opposition control (see for instance~\cite{opposition2,comparison}) and for this it was used to perform a fair comparison of the results of the two control techniques. It is important to highlight the fact that the DRL approach does not target a specific flow features as in the case of opposition control. For this, it is possible to train DRL policies based on the state observed at other wall-normal location. In this case, we train the DRL agents with the state sampling height at $y^+ = 10$. We compare the results with opposition control using the same sampling and the drag reduction performance with sampling at $y^+ = 15$ reported in~\cite{Guastoni2023}.
%Nonetheless, opposition control was tested using other wall-normal sampling locations} 
% The sampling height used in the numerical experiments is $y^+ = 15$, where $y^+$ indicated the inner-scaled coordinates using the viscous length $\ell^{*} = \nu/u_{\tau}$.
 % every agent receives as observation the velocity fluctuations in the streamwise and wall-normal directions at a given distance from the wall, above the area where it is selecting the actuation. We measure the sampling height using the \textit{inner-scaled} coordinate $y^+ = y u_{\tau}/\nu = y/\ell^{*} \in [0,Re_{\tau}]$, where $\ell^{*}$ is also termed viscous length.
% Based on these observations, the policy of the agent determines the intensity of the actuation within a prescribed range.
% Following the previous studies on opposition control~\citep{opposition2}, we consider a sampling plane at $y^+=15$. 
% %, between $-u_{\tau}$ and $u_{\tau}$.
% The policy is then updated based on the 
The reward is defined as the percentage variation of the wall-shear stress with respect to the uncontrolled flow. All the agents receive the wall-shear stress averaged on the entire wall.

% We perform the learning process using deep deterministic policy gradient (DDPG) algorithm in a small domain called \textit{minimal channel}~\citep{jimenez-minimal}. The flow dynamics in this computational domain is simpler while maintaining a good agreement in the low-order statistics with larger domains and experimental data.
% Once the learning is performed, the policy can be applied without modifications to a larger domain, with a more complete representation of all the physical features that characterize turbulent flows. Such policy transfer is possible thanks to the local nature of the control.

\subsection{Interface between the DRL framework and the numerical solver}
\begin{figure}[t!]
\begin{center}
% \begin{overpic}[width=0.7\textwidth,grid,tics=10]{channel_flow_domain_nolabels.eps}
\begin{overpic}[width=0.47\textwidth]{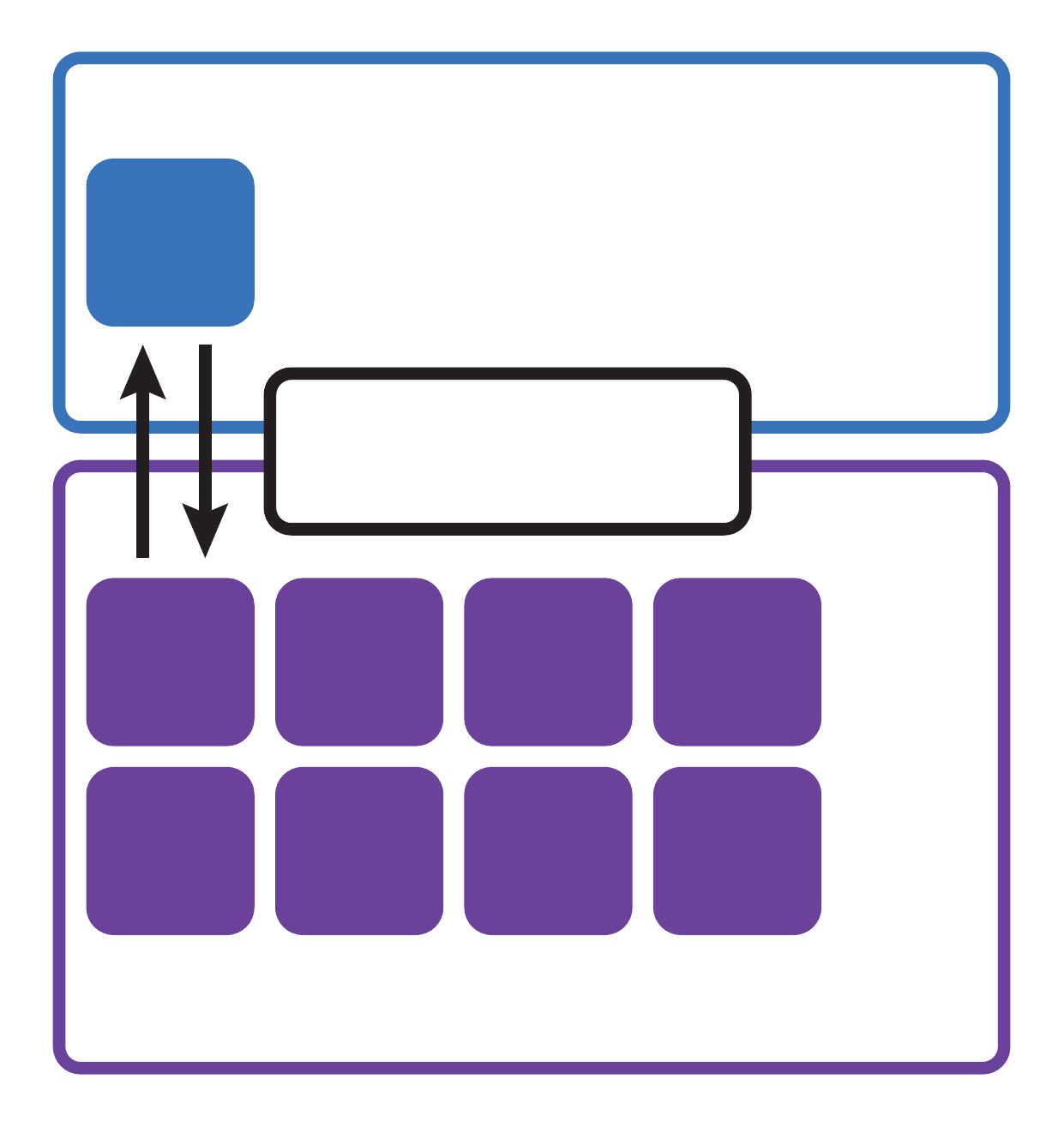}%[grid,tics=10]
 \put (35,88) {\Large \textcolor[HTML]{2277B4}{\textbf{\texttt{MPI\_COMM\_WORLD}}}}
 \put (35,8.5) {\Large \textcolor[HTML]{735096}{\textbf{\texttt{MPI\_COMM\_WORLD}}}}
 \put (27.2,58) {\Large \textbf{\texttt{MPI} Intercom}}
 \put (12.5,75.5) {\huge \textcolor{white}{$\mathbf{0}$}}
 
 \put (12.5,38) {\huge \textcolor{white}{$\mathbf{0}$}}
 \put (29.5,38) {\huge \textcolor{white}{$\mathbf{1}$}}
 \put (46,38) {\huge \textcolor{white}{$\mathbf{2}$}}
 \put (62.5,38) {\huge \textcolor{white}{$\mathbf{3}$}}
 
 \put (12.5,21.5) {\huge \textcolor{white}{$\mathbf{4}$}}
 \put (29.5,21.5) {\huge \textcolor{white}{$\mathbf{5}$}}
 \put (46,21.5) {\huge \textcolor{white}{$\mathbf{6}$}}
 \put (62.5,21.5) {\huge \textcolor{white}{$\mathbf{7}$}}
 % \put (35,84) {\Large \textcolor[HTML]{2277B4}{\begin{minipage}{0.3\textwidth}\textbf{\texttt{MPI\_COMM\_WORLD}}\\(python)\end{minipage}}}
 
 % \put (82,28) {\huge Agent}
 % \put (69,4) {\begin{minipage}{0.3\textwidth}reward\\ $r_t = 1-\tau_w/\tau_{w,\mathrm{uncontrolled}}$\end{minipage}}
 % \put (26,4) {state $\mathbf{s}_t=(u',v')$}
 % \put (77,48) {action $a_t=v_\mathrm{wall}$}
 % \put (8,18) {$y,v$}
 % \put (11,9) {$x,u$}
 % \put (1.5,9) {$z,w$}
\end{overpic}
\end{center}
\caption{\label{comm} Communication diagram between the DRL framework in python (blue) and the numerical solver in Fortran (violet). The two \texttt{MPI\_COMM\_WORLD}s exchange information through the \texttt{MPI} Intercom.
%Overview of our multi-agent DRL approach to drag reduction. The simulation domain is shown on the left. The agents are organized in a grid $N_{\rm{CTRLx}} \times N_{\rm{CTRLz}}$. Each agents observes the velocity fluctuations in the streamwise ($u'$) and wall-normal ($v'$) direction. The reward is the percentage variation of the wall-shear stress $\tau_w$. Based on the state, each agent acts by imposing a wall-normal velocity $v$ at the wall.
}
\end{figure}
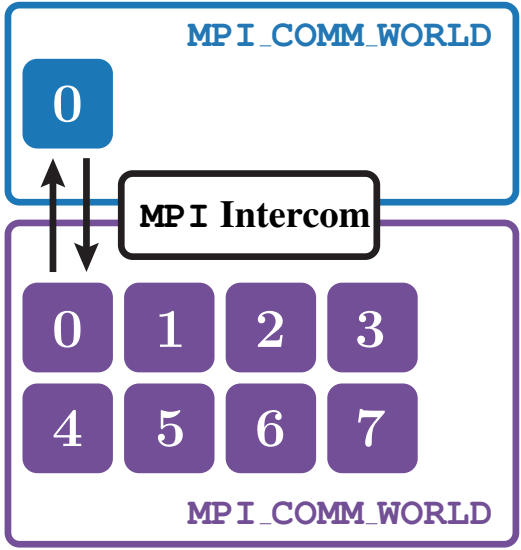

DRL algorithms require the communication between the agent and the environment at every interaction. In this case, this implies communicating states, actions, rewards between the reinforcement learning framework (in python) and the solver in Fortran. The execution of the two process is synchronous, meaning that each side of the DRL-CFD coupled system needs to periodically wait for the other before continuing. One possible approach to implement such a coupling is to let the main python program call the solver as a child process, which simulates the flow between two actuations before returning the result to the main program. %let the main program run individual step of the solver as child processes, provided that the solver allows it. 
Note that in this case the initialization steps of the solver are performed at every solver call, making the approach less efficient.
Another possibility to let the programs communicate is that one of the two programs writes the variables to disk, to be read by the other. However, this approach can significantly slow down the computation if the amount of data to be written and read is significant. 
In our approach, the fluid solver is spawned as a child process of the main learning process only once per episode (which includes hundreds or thousands of actuations) and a MPI intercomm is created between the solver and the Python main program, as pictured in figure~\ref{comm}. Importantly, by spawning a new process, a new separate communication space (\textit{i.e.} \texttt{MPI\_COMM\_WORLD}) is created. The individual processes of the solver are then ranked as if the program was run by a user. This is a crucial point for our solver as one of the processes (\texttt{MPI\_RANK}$\,=0$) acts as main program, dealing with all the gather and scatter operations, as well as writing flow fields and statistics to disk.

After the initialization, the solver can receive different information requests from the main program. Communication requests are defined as five-character strings (namely \texttt{STATE}, \texttt{CNTRL}, \texttt{EVOLV} and \texttt{TERMN}), that define the specific set of instructions for the solver to execute during each instance where the agent interacts with the environment. 
By combining these MPI messages, it is possible to define the functions that are necessary to create a new Gym environment~(\cite{gym}).

\section{Results and discussion}
\subsection{Baseline dependency on sampling height}
\begin{figure}
\begin{center}
\includegraphics[width=.47\textwidth]{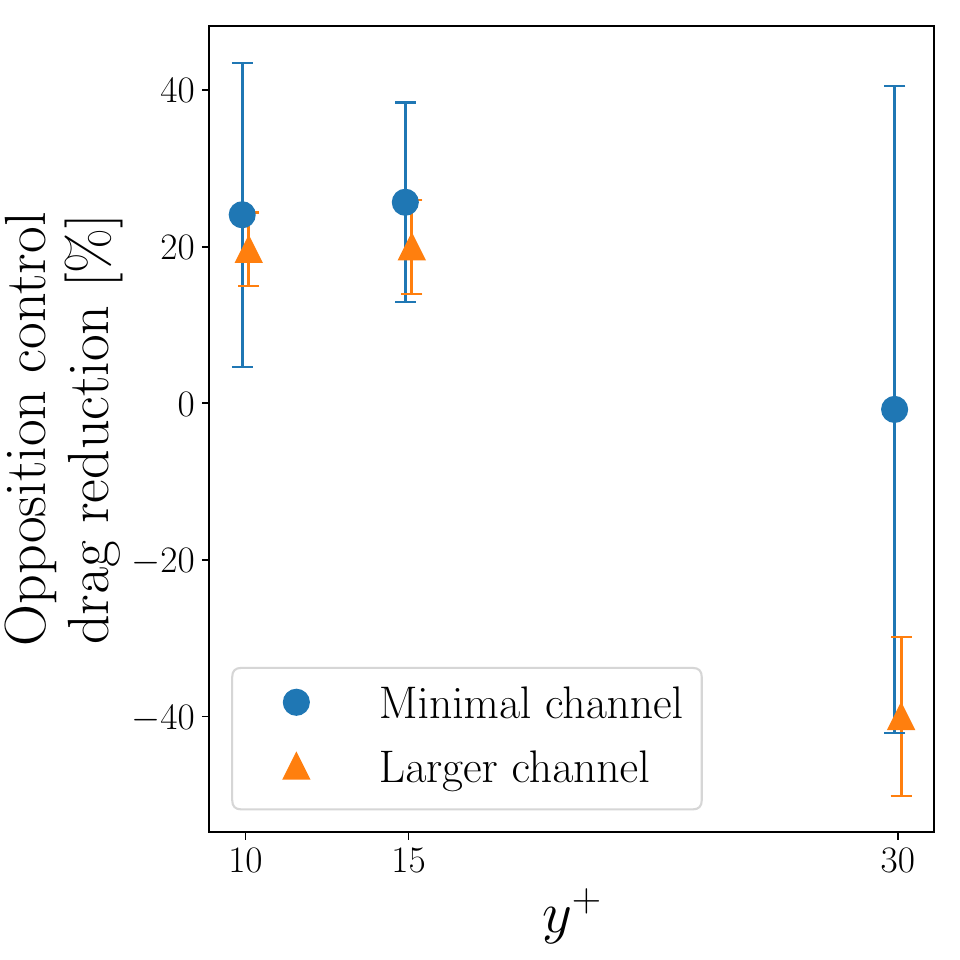}
\end{center}
\caption{\label{fig:opposition_height}  Average drag reduction with respect to the reference uncontrolled case using opposition control in the minimal channel and larger channel, using different sensing plane heights. The bars represent the standard deviation of the drag reduction with respect to the initial conditions. A small horizontal offset is introduced to enhance the readability of the results.}
\end{figure}

Opposition control is a reactive control technique that aims to suppress the wall-normal-velocity fluctuations close to the wall. This hinders the development of streamwise-parallel streaks in the flow. In this approach, the choice of the sensing plane plays an important role, helping identifying the streaky flow structures and oppose their motion. 
The actuation at the wall $v_\mathrm{wall}$ is computed as:
\begin{equation}
    v_\mathrm{wall}(x,z,t) = -\alpha\big[v(x,y_s,z,t)- \langle v(x,y_s,z,t) \rangle \big],    
\end{equation}
where $y_s$ indicates the sensing plane and $\langle v(x,y_s,z,t) \rangle$ is the spatial mean of the wall-normal velocity field. $\alpha>0$ is a scaling parameter which is fixed in the spatial directions and time. The second term of the right-hand side of the equation ensures zero-net-mass-flux within the domain. % its typical value is $\alpha=1$, independently from the height of the sampled velocity plane.
We consider $\alpha=1$, independently from the height of the sampled velocity plane. Note, however, that we limit the actuation in the range $[-u_{\tau},u_{\tau}]$ for $y^+=30$, with $u_{\tau}$ indicating the uncontrolled friction velocity. Figure~\ref{fig:opposition_height} shows the average drag reduction with the sensing plane located at different wall-normal locations. The results at $y^+ = 10$ and $y^+ = 15$ correctly reproduce the findings from previous studies in the literature. The differences between the two sampling height become less pronounced when the larger channel is considered. As we move farther away from the wall, opposition control is not able to reduce the drag because the flow structures cannot be identified based only on the information at $y^+=30$. In the minimal channel, this results in a limited effect of the control, in the larger channel we have a significant drag increase. The application of opposition control at even larger wall-normal distances has been studied~(\cite{guseva_jimenez_2022}) and drag increase is reported also in this case. In the remainder of the proceeding, we consider the two wall-normal locations that are closer to the wall. %it defines the intensity of the control applied at the wall. If the 

\subsection{DRL results}
\begin{figure}
\begin{center}
\includegraphics[width=.47\textwidth]{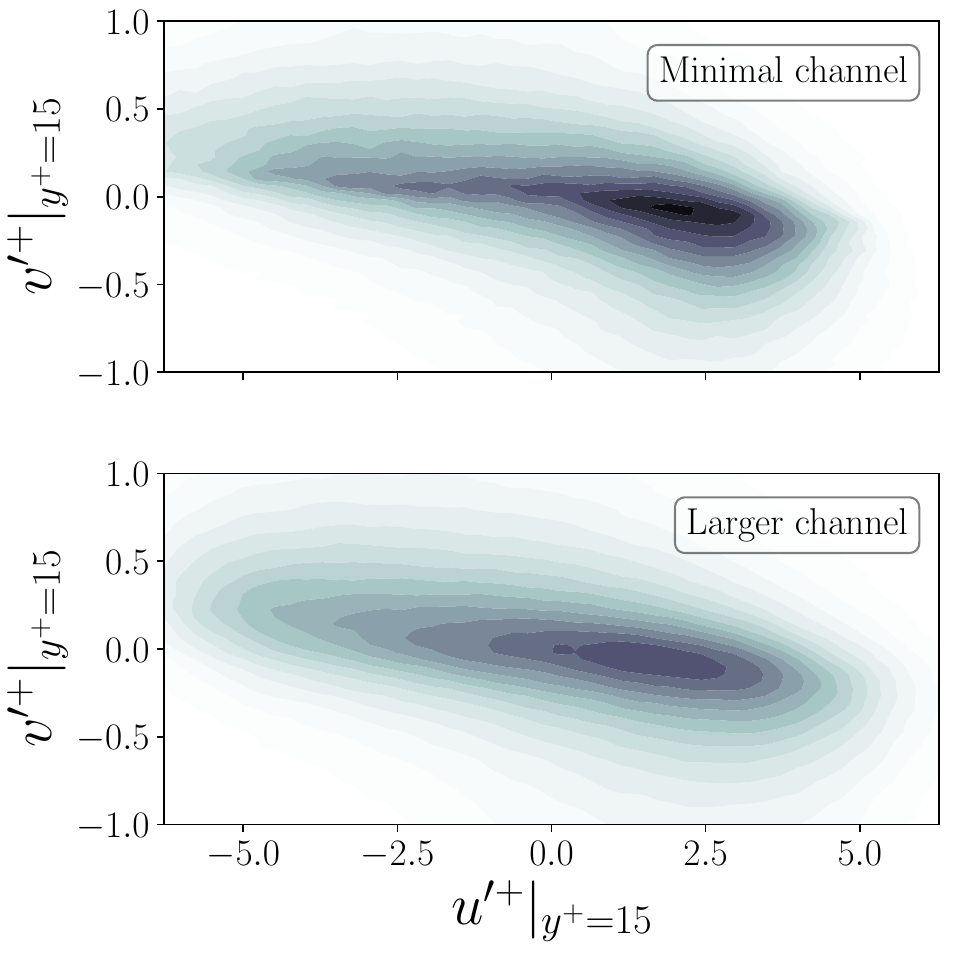}
\end{center}
\caption{\label{fig:quadrant} Distribution in the uncontrolled case of the inner-scaled velocity-fluctuation components in the streamwise ($u$) and wall-normal ($v$) directions at $y^+=15$, in the minimal channel (top) and larger one (bottom).}
\end{figure}

We perform the learning process using deep deterministic policy gradient (DDPG) algorithm~(\cite{ddpg}) in the minimal channel.
As described in more detail in~\cite{Guastoni2023}, the policy learnt using deep reinforcement learning is consistently providing an average drag reduction that is higher than the one obtained with opposition control. After the initial transient ($t^+>500$) the DDPG policy provides 43\% drag reduction, while opposition control is limited to 26\%. Furthermore, having a local and translational invariant policy allows us to apply it to a different domain size with no modifications. This way, we are able to obtain 30\% drag reduction in the larger channel, whereas opposition control is limited to 20\%. 
Here we apply the same approach to study the DRL policies obtained by observing the state at a different wall-normal location. In particular, we train a DRL agent using the state at $y^+=10$. The learnt policy is tested on 6 different initial conditions, both in the minimal and larger channel. The drag reduction achieved in the two numerical domains is $35\%$ and $28.8\%$, respectively. The DRL policy is consistently better than opposition control (whose drag-reduction figures are shown in figure~\ref{fig:opposition_height}) also with this sensing height. While the reduction in the minimal channel is significantly smaller than the one obtained with the state observed at $y^+=15$, the difference in the larger channel is limited to two percentage points. This highlights the importance of testing the DRL policy also in the larger channel. 
The performance reduction may depend on the fact that in the larger channel there are some physical features that cannot be experienced by the agent in the minimal channel. The limitations of the minimal channel in the simulation of the flow physics are reported in~\cite{jimenez-minimal}, however the DRL agents only have access to a portion of the flow field. Figure~\ref{fig:quadrant} compares the distribution at $y^+=15$ of the streamwise and wall-normal fluctuations of the velocity in the uncontrolled case. While the overall shape of the distribution is the same, the minimal channel distribution has higher density in Q4 ($u'>0$, $v'<0$) and also negative streamwise fluctuations associated with very small wall-normal fluctuations, between Q2 ($u'<0$, $v'>0$) and Q3 ($u'<0$, $v'<0$).
A different input distribution determines a different wall-normal velocity distribution at the wall for control.

\begin{figure*}
\begin{center}
\includegraphics[width=\textwidth]{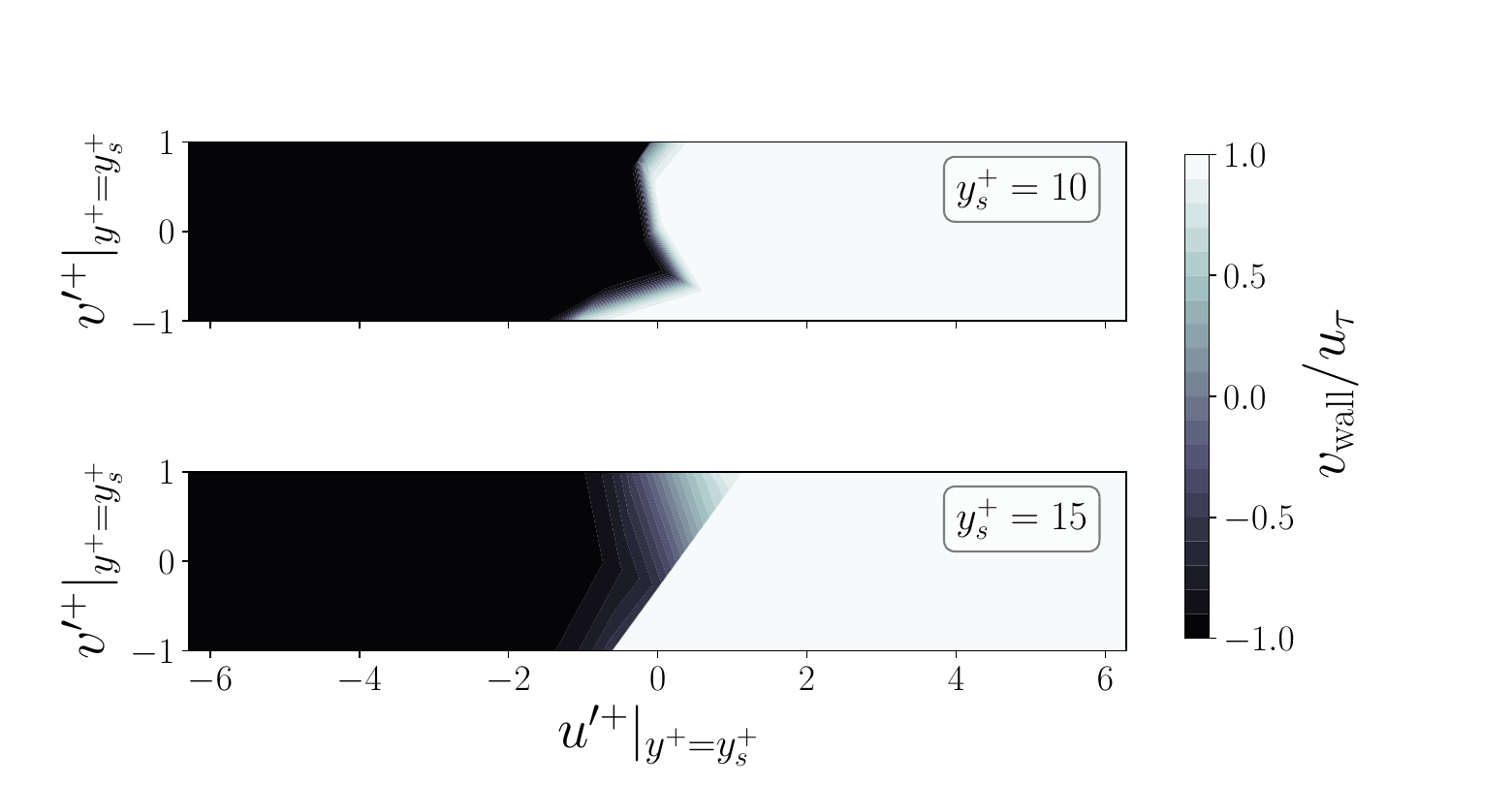}
\end{center}
\caption{\label{fig:in_out} Input-output maps of the DRL control policies learnt in the minimal channel, using the state observed at $y^+ = 10$ (top) and $y^+ = 15$ (bottom)}
\end{figure*}

DDPG is a learning algorithm that provides a deterministic policy, meaning that it is possible to visualize the input-output relation, as shown by~\cite{sonoda}. Figure~\ref{fig:in_out} compares the policies learnt in the minimal channel using the state at $y^+=10$ and $y^+=15$. The overall behaviour is similar in both cases: the policy is mostly sensitive to the streamwise fluctuations, while the variation of the wall-normal fluctuations has a limited effect. Suction is applied for negative streamwise velocity fluctuations, with blowing is used for positive ones.
% The control strategy learnt with deep reinforcement learning have a profound effect on the flow physics and the velocity-fluctuations distribution. 

\section{Conclusions}
In this work, we implemented the coupling of a high-performance fluid-dynamics solver and a (deep) multi-agent reinforcement-learning framework using dynamic process management in MPI, with the objective to discover novel drag-reduction strategies.
Our retrofitting approach using MPI can be potentially applied to other existing HPC codes in Fortran/C++, not necessarily used for fluid-dynamics simulations. This way, other computationally-intensive scientific domains can explore the potential of DRL in other tasks without the need to rewrite their codebase.

We assessed the effect of the sampling-plane height in both opposition control and in the learning process using DRL.
By using the DDPG algorithm, we are able to design a control policy that outperforms existing strategies like opposition control both at $y^+=10$ and $y^+=15$. As we move farther away from the wall, opposition control struggles to provide drag reduction because its control strategy is based on the suppression of the near-wall turbulent structures that are difficult to observe farther from the wall. At $y^+=30$, the application of opposition control yields a drag increase. DRL has the potential to learn different strategies for drag reduction, using information at a different wall-normal location. The comparison of the policies learnt observing the state at $y^+=10$ and $y^+=15$ represents the first step towards the complete assessment of the DRL potential.

\Acknowledgments

The code supporting the findings in this paper including the environment implementation is available in the repository on \burlalt{Github}{https://github.com/KTH-FlowAI/MARL-drag-reduction-in-wall-bounded-flows}\footnote{\url{https://github.com/KTH-FlowAI/MARL-drag-reduction-in-wall-bounded-flows}}.
The authors acknowledge the Swedish National Infrastructure for Computing (SNIC) for providing the computational resources by PDC, used to carry out the numerical simulations. This work is supported by the founding provided by the ERC grant no.~"2021-CoG-101043998, DEEPCONTROL".%, the Swedish e-Science Research Centre (SeRC) and the Knut and Alice Wallenberg (KAW) Foundation.

%%% References %%%

% \begin{References}
% \item Haecheon Choi, Parviz Moin, and John Kim. Active turbulence control for drag reduction in wall-bounded flows. Journal of Fluid Mechanics, 262:75–110, 1994.
% \item Edward P. Hammond, Thomas R. Bewley, and Parviz Moin. Observed mechanisms for turbulence attenuation and enhancement in opposition-controlled wall-bounded flows. Physics of Fluids, 10(9):2421–2423, 1998. doi: 10.1063/1.869759
% % \item Jones A.B., Jordan, D.L and March, F.P. (2002), Linear aspects of transition, {\it J. Fluid Mech.}, Vol. 34, pp. 123-153. 
% % \item Smith, W.S. and Martin, W.P. (1997), Leading edge paper on control mechanism, AIAA Paper 97-1234.

\begin{References}
\printbibliography[heading=none]
\end{References}

\end{document}